\documentclass[preprint,prl,aps,showpacs]{revtex4}

\usepackage{bm,here}
\usepackage{graphicx}
\usepackage{amsmath,amsxtra,amsfonts,amssymb,amstext,latexsym}

\usepackage{times}
\usepackage{mathptm}

\newcommand{\p}{\partial}
\newcommand{\wh}{\widehat}
\newcommand{\wt}{\widetilde}
\newcommand{\eps}{\epsilon}

\newcommand{\vek}[1]{{\mathbf#1}}

\newcommand{\uvek}[1]{{\wh{\mathbf#1}}}

\newcommand{\nablan}{{\nabla_{\perp}}}

\newcommand{\etal}{\textit{et al}.}

\newcommand{\ibid}{\textit{ibid}.}
\newcommand{\pdf}{{PDF}}
\newcommand{\sol}{{SOL}}
\newcommand{\lcfs}{{LCFS}}

\begin{document}


\title{Numerical Simulations of Intermittent Transport in Scrape-Off Layer Plasmas}
\author{O.~E.~Garcia}
\author{V.~Naulin}
\author{A.~H.~Nielsen}
\author{J.~Juul Rasmussen}
\affiliation{Association EURATOM-Ris{\o} National Laboratory,
OFD-128 Ris{\o}, DK-4000 Roskilde, Denmark}
\date{\today}

\begin{abstract}
Two-dimensional fluid simulations of interchange turbulence for geometry
and parameters relevant for the scrape-off layer of confined plasmas are
presented. We observe bursty ejection
of particles and heat from the bulk plasma in the form of blobs. These
structures propagate far into the scrape-off layer where they are lost
due to transport along open magnetic field lines. From single-point
recordings it is shown that the blobs have asymmetric conditional wave
forms and lead to positively skewed and flat probability distribution
functions. The radial propagation velocity may reach one tenth of the
sound speed. These results are in excellent agreement with recent
experimental measurements.
\end{abstract}

\maketitle

Recently, several experimental investigations have revealed a
strongly intermittent nature of particle and heat transport in the
scrape-off layer (\sol) of magnetized plasmas~\cite{ct:boedo,ct:antar,%
ct:terry}. There are strong indications that this is caused by localized
structures in the form of plasma ``blobs'' propagating radially far into
the \sol. It has been suggested that this is due to the dipolar vorticity
field caused by vertical guiding center motions in plasmas in non-uniform
magnetic fields~\cite{ct:krash}, reflecting the
compressibility of the diamagnetic current.
An outstanding challenge is to give a self-consistent description of the
emergence and evolution of such structures, which also capture their
statistical properties. Here an attempt towards this goal is presented,
yielding favorable agreement with experimental measurements. This is
achieved by focusing on the collective two-dimensional dynamics
perpendicular to the magnetic field while using a simplified description
of particle and heat losses to limiters or end plates along open field
lines.

Some of the most prominent features of experimental single-point
measurements are the presence of asymmetric conditional wave forms
as well as strongly skewed and flat probability distribution functions
(\pdf's) of the density and temperature signals~\cite{ct:boedo,ct:antar,%
ct:terry}. Simple interpretations as
well as advanced imaging techniques give a picture of field-aligned
blobs or filaments propagating out of the bulk plasma with radial
velocities up to one tenth of the sound speed. These highly non-linear
thermal structures have amplitudes which significantly exceed the
back-ground levels. The associated intermittent transport may have
severe consequences for magnetic confinement experiments by producing
large heat bursts on plasma facing components. A change of sign in the
asymmetry of the fluctuation time series PDF close to the
last closed flux surface (\lcfs) indicates that the structures are
generated by a ``flapping'' of the edge pressure gradient, ejecting
blobs of excess particles and heat out of the bulk plasma~\cite{ct:boedo}.
The blob structures thus seems to be generated close to the \lcfs\ and
subsequently propagate far into the \sol, where they are subject to
strong damping due to parallel losses. This separation of driving and
damping regions in configuration space has been discarded in several
previous studies of \sol\ turbulence~\cite{ct:sarazin,ct:benkadda}.

In this Letter we present a novel model for interchange turbulence
in slab geometry and numerical solutions in qualitative agreement
with experimental measurements. The model geometry comprises
distinct production and loss regions, corresponding to the edge and
\sol\ of magnetized plasmas. The separation of these two regions
defines an effective \lcfs, though we do not include magnetic shear in
our model. In the edge region, strong pressure gradients maintain a
state of turbulent convection. It is demonstrated that a self-regulation
mechanism involving differential rotation leads to a repetitive
expulsion of hot plasma into the \sol, resulting in asymmetric
conditional wave forms, strongly non-Gaussian probability
distributions and significant cross-field transport by localized objects.

Assuming cold ions and neglecting electron inertia effects, a three-field
model may be derived for quasi-neutral electrostatic perturbations of the
full particle density $n$, electric potential $\phi$ and electron temperature
$T$. Using the Bohm normalization and slab coordinates with $\uvek{z}$
along the magnetic field we obtain~\cite{ct:model},
\begin{gather*}
\left( \frac{\p}{\p t} + \uvek{z}\times\nabla\phi\cdot\nabla \right) \Omega -
\mathcal{C}(p) =  \nu_{\Omega} \nabla^2 \Omega - \sigma_{\Omega} \Omega , \\
\frac{dn}{dt} + n\mathcal{C}(\phi) - \mathcal{C}(nT) =
\nu_n \nabla^2 n - \sigma_n ( n-1 ) + S_n  , \\
\frac{dT}{dt} + \frac{2T}{3}\,\mathcal{C}(\phi) -
\frac{7T}{3}\,\mathcal{C} (T) - \frac{2T^2}{3n}\,\mathcal{C}(n)
= \nu_{T} \nabla^2 T - \sigma_{T}( T-1 ) + S_{T} ,
%
\end{gather*}
where time is normalized by the ion gyration period, $1/\omega_{ci}$, and
spatial scales by the hybrid gyration radius, $\rho_s=c_s/\omega_{ci}$.
The particle density $n$ and temperature $T$ are normalized to fixed
characteristic values at the outer wall. We further define the
two-dimensional advective derivative, the magnetic field curvature
operator and the toroidal magnetic field, respectively, by
\[
\frac{d}{dt} = \frac{\p}{\p t} +
\frac{1}{B}\,\uvek{z}\times\nabla\phi\cdot\nabla , \quad
\mathcal{C} = - \zeta\,\frac{\p}{\p y}  ,  \quad
B = \frac{1}{1 + \eps + \zeta x} .
\]
The vorticity is given by $\Omega=\nabla_{\perp}^2\phi$, the inverse aspect
ratio $\eps=a/R_0$ and $\zeta=\rho_s/R_0$ where $a$ and $R_0$ are the minor
and major radius of the device, respectively. The terms on the right hand side
of the model equations describe external sources $S$, parallel losses along
open field lines through the damping rates $\sigma$, and collisional diffusion
with coefficients $\nu$. The geometry and boundary conditions are sketched
in Fig.~\ref{fig:geometry}.

In the absence of external forcing and dissipative processes the model
equations non-linearly conserves the global energy
\[
E = \int d\vek{x}\:\left[ \frac{1}{2}\left( \nablan\phi \right)^2 +
\frac{3}{2}\,nT \right] ,
\]
where the integral extends over the whole plasma layer under consideration.
Thus, the curvature terms due to magnetic field inhomogeneity correctly
yield a conservative energy transfer from the confined heat to the convective
motions. We further define the kinetic energy of the fluctuating and mean
components of the flows,
\begin{equation} \label{kinenergy}
K = \int d\vek{x}\:\frac{1}{2}\left( \nabla_{\perp}\wt{\phi} \right) ,
\qquad
U = \int d\vek{x}\:\frac{1}{2}\,v_0^ 2 ,
\end{equation}
where the zero index denotes an average over the periodic direction $y$
and the spatial fluctuation about this mean is indicated by a tilde. The
linearly damped mean flows, $v_0=\p\phi_0/\p x$, does not yield any radial
convective transport and hence form a benign path for fluctuation energy.
The energy transfer rates from thermal energy to the fluctuating motions,
and from the fluctuating to the mean flows, are given respectively by
\begin{equation} \label{transfer}
F_p = \int d\vek{x}\:nT\mathcal{C}(\phi) ,
\qquad
F_{v} = \int d\vek{x}\:\wt{v}_x\wt{v}_y\frac{\p v_0}{\p x} .
\end{equation}
Note that $F_p$ is essentially a measure of the domain integrated
convective thermal energy transport, while $F_v$ shows that structures
tilted such as to transport positive poloidal momentum up the gradient
of a sheared flow will sustain the flow against
collisional dissipation~\cite{ct:garcia,ct:gb,ct:naulin}.

In the following we present results from numerical simulations of the
interchange model using parameters relevant for \sol\ plasmas. The
dimensions of the simulation domain is
$L_x=2L_y=400$ and the \lcfs\ is located at $x_\text{\tiny LCFS}=100$.
The parameters used for the simulation presented here are $\eps=0.25$,
$\zeta=10^{-3}$, and the collisional diffusion $\nu=5\times10^{-3}$ is
taken to be the same for all fields. The parallel loss rate of
temperature is assumed to be five times stronger than that on the density
and vorticity, $\sigma_n=\sigma_{\Omega}=\sigma_T/5=10^{-3}/2\pi q$,
since primarily hot electrons are lost through the end sheaths~\cite{ct:s}.
The damping rates $\sigma_n$ and $\sigma_\Omega$ correspond to
losses along one connection length $2\pi R_0 q$ with the acoustic
speed $c_s$, where $q=3$ is the safety factor at the edge. Finally, the
radial line-integral of the sources $S_n$ and $S_T$ equals $0.2$,
and the shape of the sources and parallel loss coefficients shown in
Fig.~\ref{fig:geometry} are given by $\delta=16$ and $\xi=2$. For the
numerical solution we have employed an Arakawa scheme for the advective
non-linearities and a third order stiffly stable scheme for the time
integration~\cite{ct:model}. The spatial resolution is 512 and 256 grid
points in the radial and poloidal directions, respectively, and the total
time span of the simulation  is $4\times10^6$.

In Fig.~\ref{fig:confine} we show the typical evolution of the
particle confinement $P$ and heat confinement $H$ in the edge and
\sol\ regions, defined by
\[
P_\text{edge} = \int_{0}^{x_\text{\tiny LCFS}} dx\;n_0(x,t) , \qquad
P_\text{\tiny SOL} = \int_{x_\text{\tiny LCFS}}^{L_x} dx\;n_0(x,t) ,
\]
and similarly for the heat confinement $H$. From the figure we observe
that plasma and heat gradually builds up in the edge at the same time as
it is decaying in the \sol\ region. This is repetitively interrupted
by rapid changes in which plasma and heat is lost from the edge to the
\sol. More than 20\% of the edge plasma may be lost during individual
bursts. Also note from the figure that the normalized heat confinement
is much less than the particle confinement due to the larger loss rate
in the \sol\ region of the former. Further insight is revealed by
Fig.~\ref{fig:kinetic} which shows the evolution of the kinetic
energy contained by the mean and fluctuating motions, confer
Eq.~\eqref{kinenergy}, as well as the collective energy transfer
terms defined in Eq.~\eqref{transfer}. From the figure we observe that
the convective energy and thermal transport appears as bursts during
which particles and heat are lost from the edge into the \sol\ region.
As discussed in Refs.~\onlinecite{ct:garcia,ct:gb,ct:naulin}, this
global dynamics is caused by a self-regulation mechanism in which
kinetic energy is transfered from the fluctuating to the mean components
of the flows, and subsequently damped by collisional dissipation. As the
thermal confinement is allowed to vary, this results in characteristic
sawtooth oscillations~\cite{ct:garcia}. Note, however, the clear
demonstration in Figs.~\ref{fig:confine} and~\ref{fig:kinetic}
that thermal energy is ejected in a bursty manner from the edge and into
the \sol\ region, where it is eventually lost by transport along open
field lines.

The time-averaged profiles of the plasma density and the mean poloidal
flows are shown in Fig.~\ref{fig:profile}. Parallel losses in the \sol\
region result in average profiles peaked inside the \lcfs, and weakly
decaying throughout the \sol. Also shown in Fig.~\ref{fig:profile} are
typical instantaneous profiles during a quiet period ($t_\text{quiet}$)
and during a burst ($t_\text{burst}$). We observe significant deviations
from the average profiles, with a more peaked density profile in the edge
region during quiet phases. During bursts there is a substantial increase
of the density profile in the \sol\ region due to the convective plasma
transport. The temperature profiles have a similar structure but with lower
amplitudes due to the larger loss rate. The self-sustained poloidal flow
profiles are strongly sheared in the edge region, and have larger
amplitudes during the the strong fluctuation period. This figure
clearly indicates that fluctuations are driven in the strong pressure
gradients in the edge region.

The statistics of single-point recordings at different radial positions
$P_i$ indicated in Fig.~\ref{fig:geometry} completely agree 
with experimental measurements. In Fig.~\ref{fig:pdf} we present the
probability distribution functions of the density signals taken from
a long-run simulation containing more than a hundred burst events.
The flat and strongly skewed distributions indicate the high probability
of large positive fluctuations corresponding to blobs of excess plasma.
The skewness and flatness factors take values up to 3 and 15,
respectively, except for the outermost radial position ($P_7$) where
very few structures arrive. At the other points the \pdf's have similar
structure with a pronounced exponential tail towards large values, a
characteristic feature of turbulent convection in the presence of sheared
flows~\cite{ct:garcia,ct:gb}.

The conditionally averaged temporal wave forms calculated from the same
signals, using the trigger condition $n>4n_\text{rms}$ at each individual
point, are presented in Fig.~\ref{fig:conditional}. An asymmetric wave
form with a sharp rise and a relatively slow decay, as observed in
experimental measurements, is clearly seen~\cite{ct:boedo,ct:antar,%
ct:terry}. The maximum density excursions significantly exceed the
background level, and decay rapidly as the structures propagate through
the \sol.  The number of realizations for
the conditional averaging decreases gradually from 125 at the innermost
probe to 2 at the outermost one. Using a negative amplitude for the
conditional averaging results in very few realizations, again showing
the presence of blob-like structures. From two-dimensional animations
and simple statistical correlation measurements we further find that
the radial propagation velocity of these structures is typically about
one tenth of the sound speed but with a large statistical variance.
This is in excellent agreement with experimental
measurements~\cite{ct:boedo,ct:antar,ct:terry}.

In Fig.~\ref{fig:structure} we show the spatial structure of the density,
temperature, vorticity and electrostatic potential during a quiet
period and during a burst. These correspond to the same times as the
instantaneous profiles shown in Fig.~\ref{fig:profile}. In the quiet
period there are only weak spatial fluctuations with the plasma and heat
well confined within the \lcfs. During bursts we observe strong structures
in all fields, which have propagated far into the \sol. Notice again the
much weaker perturbation in the temperature field as compared to the
density due to the difference in the parallel loss rates. Moreover,
while blob-like structures are observed for the density and temperature
fields, the vorticity displays a roughly dipolar structure as expected
from theory and experimentally measured~\cite{ct:krash,ct:boedo}.

In this Letter we have proposed a new model for interchange turbulence
and demonstrated that its numerical solutions are in good agreement with
that reported from experimental investigations of \sol\ turbulence and
transient transport events~\cite{ct:boedo,ct:antar,ct:terry}. An
important feature of our model is the spatial
separation between forcing and damping regions which has not been accounted
for in previous studies~\cite{ct:sarazin,ct:benkadda}. 
While our model does not in detail describe the transition region between
closed and open magnetic field lines, its solution clearly demonstrates a
sound mechanism for the origin and nature of intermittent transport in the
\sol\ of magnetized plasmas. Our work verifies the present experimental
working hypothesis in terms of field-aligned blob-like structures
propagating far into the scrape-off layer, which was recently questioned
by Sarazin \etal~\cite{ct:sarazin}. The blob-like transport results
in strongly skewed and flat \pdf's and asymmetric conditional wave forms.
This is caused by a ``flapping'' of the edge
pressure profile, ejecting blobs of excess particles and heat out of the
bulk plasma. The excellent qualitative agreement between simulation
data and experimental observations gives strong confidence that the
two-dimensional structure in potential, vorticity, density and
temperature of the blobs reveals the actual spatial shape of these
blobs as occurring in experiments.

This work was supported by the Danish Center for Scientific Computing
through grants no.\ CPU-1101-08 and CPU-1002-17. O.~E.~Garcia has been
supported by financial subvention from the Research Council of Norway.


\newpage

\begin{figure}[H]
\caption{Geometry of the simulation domain showing the
forcing region to the left, corresponding to the edge plasma, and the
parallel loss region to the right, corresponding to the scrape-off
layer. Data time series are collected at the probe positions $P_i$.}
\label{fig:geometry}
\end{figure}

\begin{figure}[H]
\caption{Evolution of particle confinement $P$ and the heat confinement
$H$ in the edge and scrape-off layer regions, showing sawtooth oscillations.}
\label{fig:confine}
\end{figure}

\begin{figure}[H]
\centering
\caption{Evolution of the kinetic energies and the collective energy
transfer terms, showing bursty behavior in the fluctuation integrals.}
\label{fig:kinetic}
\end{figure}

\begin{figure}[H]
\centering
\caption{Time-averaged profiles of plasma particle density $n_0$ and
mean poloidal flow $v_0$, and typical profiles during a fluctuation
burst and during a quite period.}
\label{fig:profile}
\end{figure}

\begin{figure}[H]
\caption{Probability distribution functions of density measured at
seven different radial positions $P_i$ as shown in Fig.~\ref{fig:geometry}.
The vertical axis shows count numbers.}
\label{fig:pdf}
\end{figure}

\begin{figure}[H]
\caption{Conditionally averaged wave forms of the density measured at
seven different radial positions $P_i$ as shown in Fig.~\ref{fig:geometry},
using the condition $n(x_{P_i})>4n_\text{rms}$.}
\label{fig:conditional}
\end{figure}

\begin{figure}[H]
\caption{Typical spatial structure of density, temperature, vorticity
and electric potential during a quite period to the left and during a
burst to the right.}
\label{fig:structure}
\end{figure}

\newpage

\includegraphics[width=.45\textwidth]{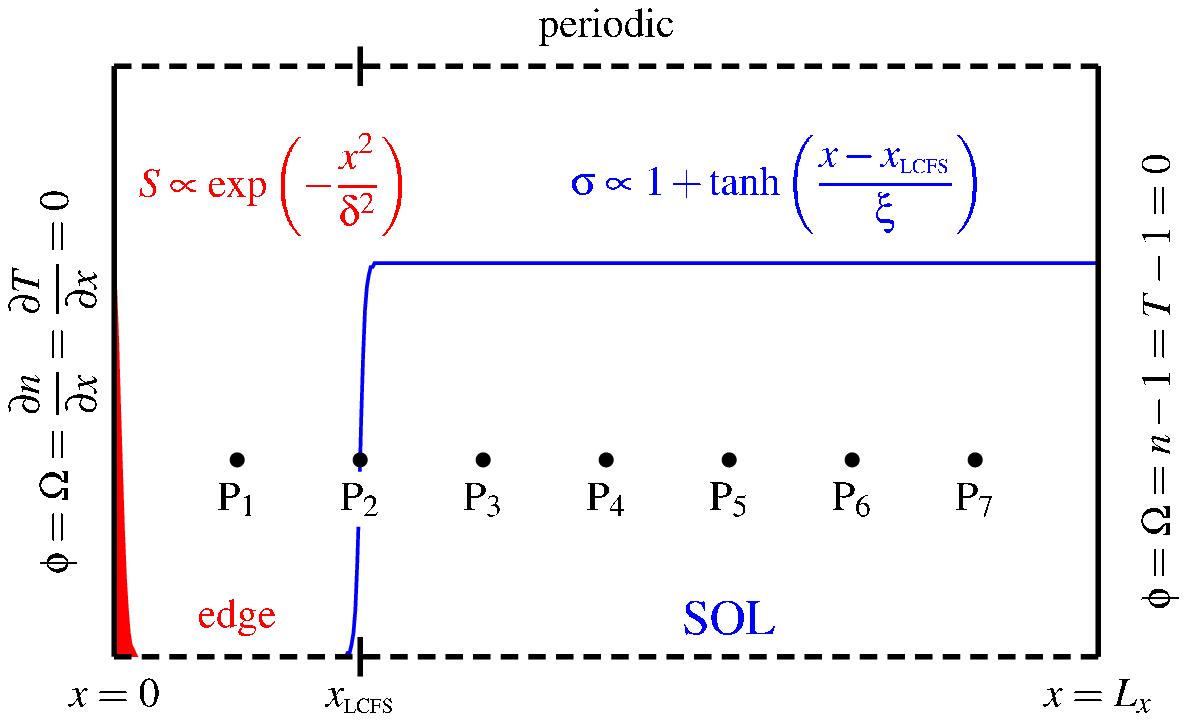}\\
{Figure 1}

\newpage

\includegraphics[width=.45\textwidth]{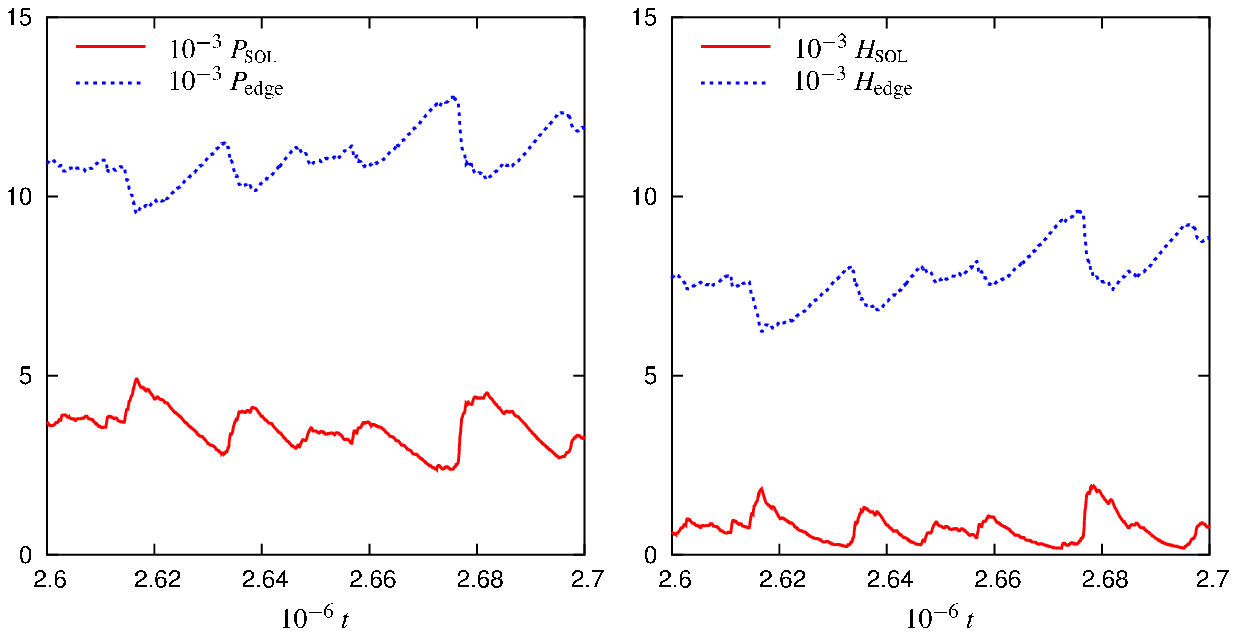}\\
{Figure 2}

\newpage

\includegraphics[width=.45\textwidth]{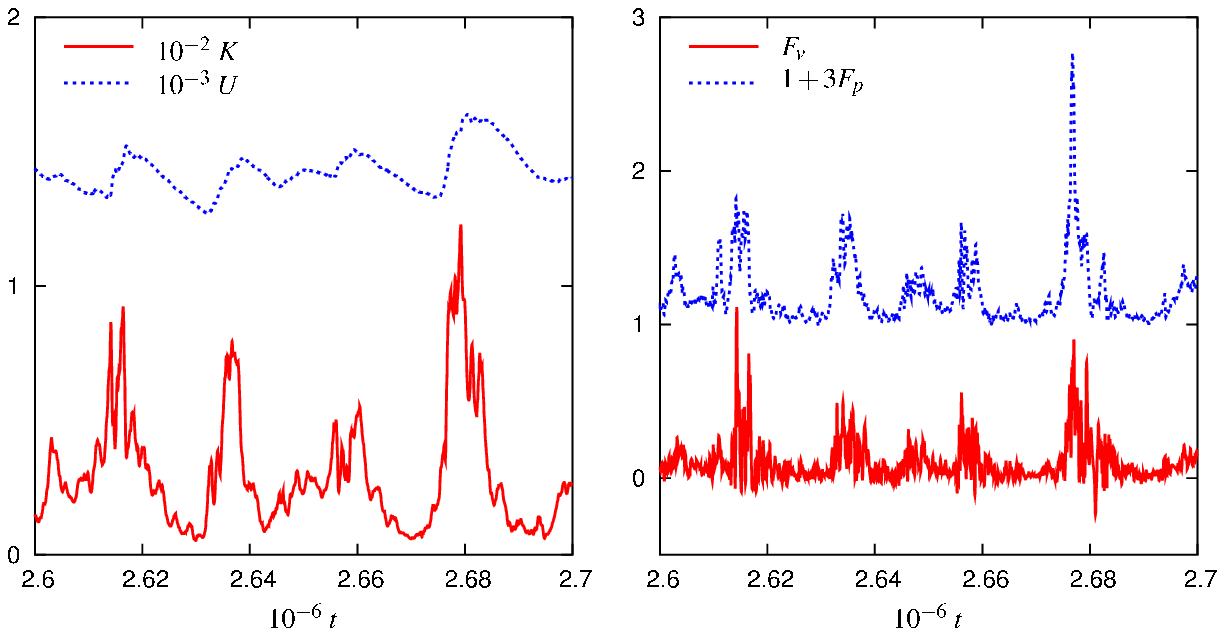}\\
{Figure 3}

\newpage

\includegraphics[width=.45\textwidth]{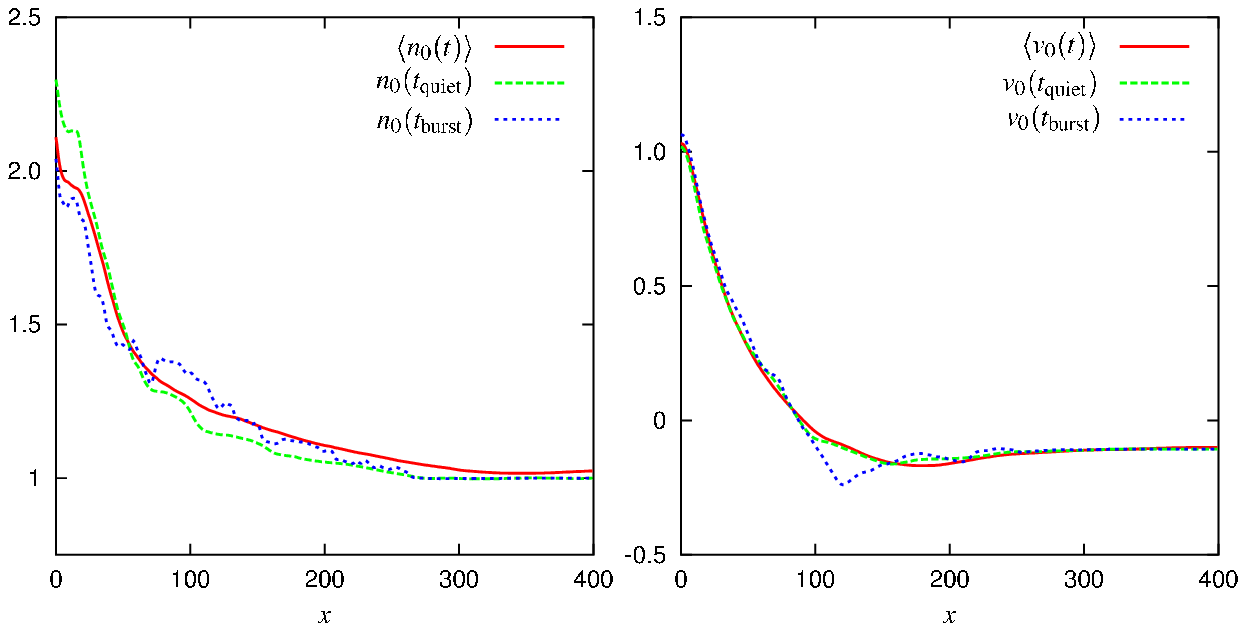}\\
{Figure 4}

\newpage

\includegraphics[width=.45\textwidth]{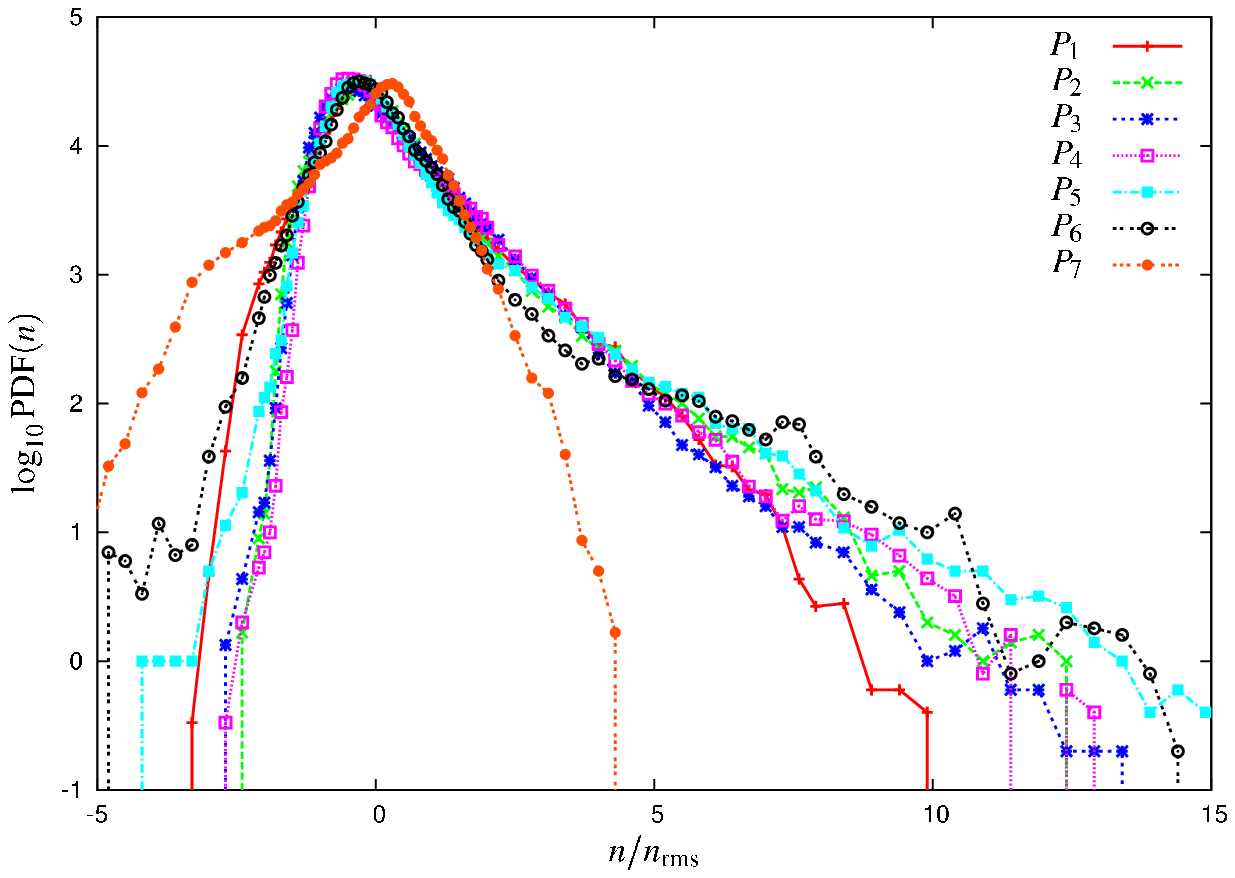}\\
{Figure 5}

\newpage

\includegraphics[width=.45\textwidth]{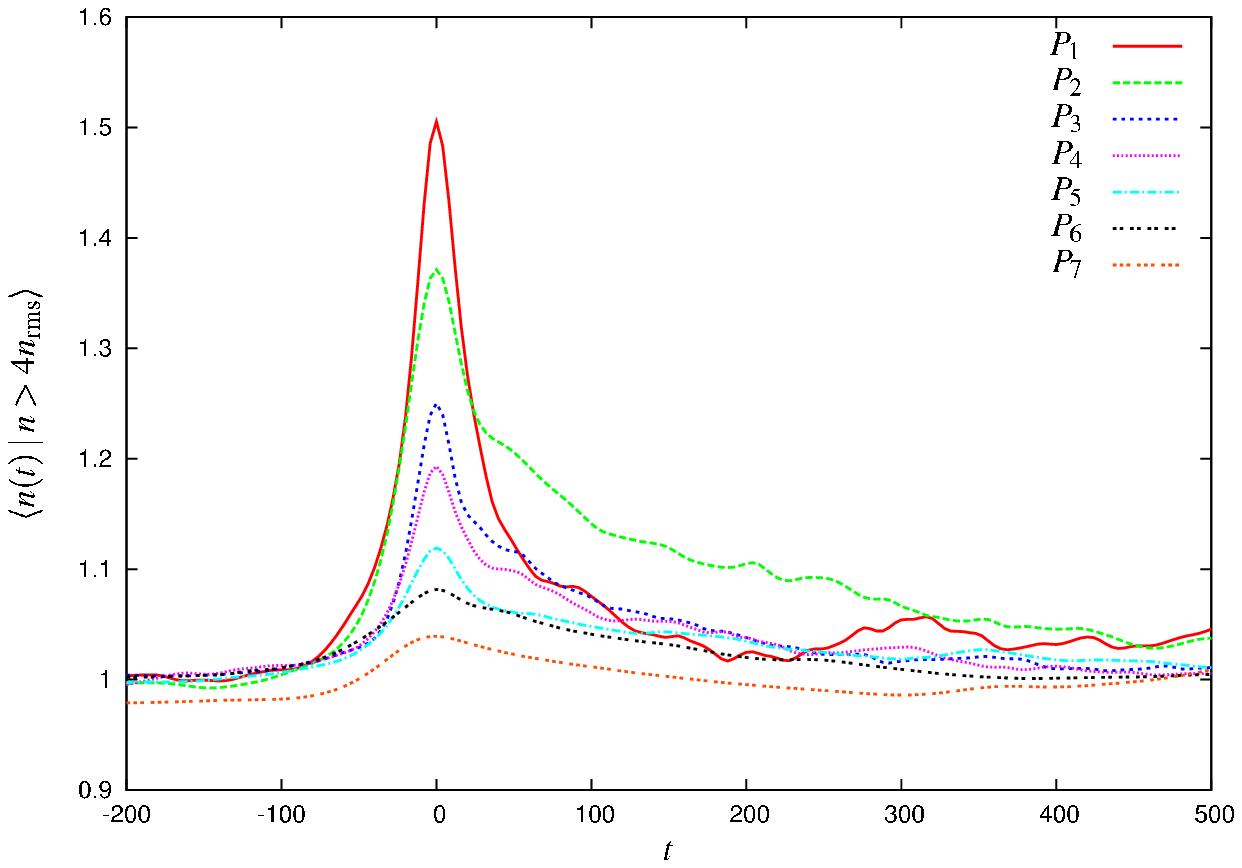}\\
{Figure 6}

\newpage

\includegraphics[width=.45\textwidth]{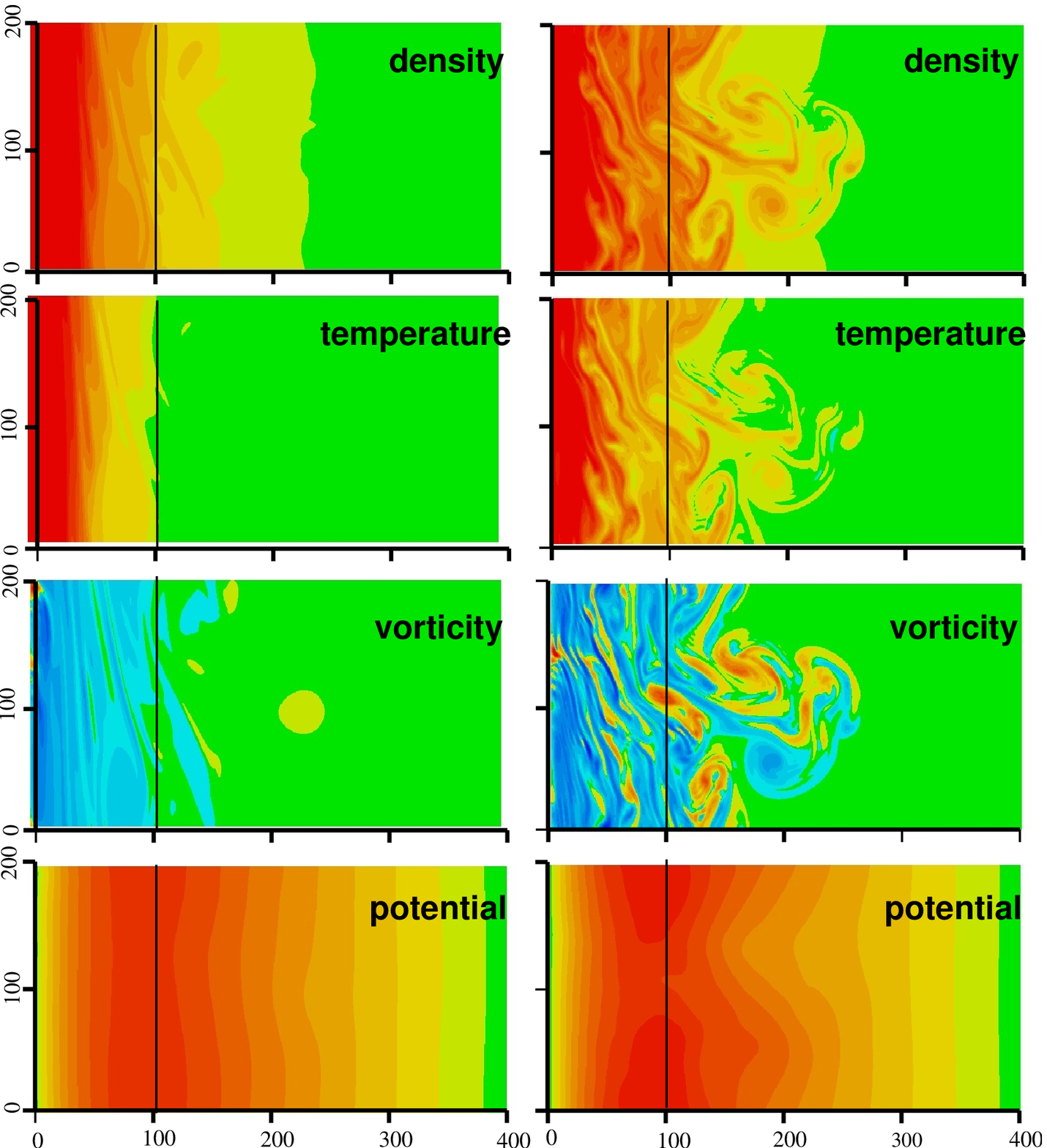}\\
{Figure 7}


\begin{thebibliography}{99}
%
\bibitem{ct:boedo}
J.~A.~Boedo \etal, J.~Nucl.\ Mater.\ {\bf 313--316}, 813 (2003);
Phys.\ Plasmas {\bf 10}, 1670 (2003);
\ibid\ {\bf 8}, 4826 (2001)
D.~L.~Rudakov \etal, Plasma Phys.\ Control.\ Fusion {\bf 44}, 717 (2002).
%
\bibitem{ct:antar}
G.~Y.~Antar \etal, Phys.\ Plasmas {\bf 10}, 419 (2003); 
\ibid\ {\bf 8}, 1612 (2001); Phys.\ Rev.\ Lett.\ {\bf 87}, 065001 (2001).
%
\bibitem{ct:terry}
J.~L.~Terry \etal, Phys.\ Plasmas {\bf 10}, 1739 (2003); 
S.~J.~Zweben \etal, \ibid\ {\bf 9}, 1981 (2002);
R.~J.~Maqueda \etal, \ibid\ {\bf 8}, 931 (2001).
%
%
\bibitem{ct:krash}
S.~I.~Krasheninnikov, Phys.\ Lett.~A {\bf 283}, 368 (2001)
D.~A.~D'Ippolito, J.~R.~Myra and S.~I.~Krasheninnikov,
Phys.\ Plasmas {\bf 9}, 222 (2002); N.~Bian, S.~Benkadda,
J.-V.~Paulsen and O.~E.~Garcia, \ibid\ {\bf 10}, 671 (2003).
%
\bibitem{ct:sarazin}
Y.~Sarazin, Ph.~Ghendrih, G.~Attuel, C.~Cl{\'e}ment, X.~Garbet,
V.~Grandgirard, M.~Ottaviani, S.~Benkadda, P.~Beyer, N.~Bian, C.~Figarella,
J.~Nucl.\ Mater.\ {\bf 313--316}, 796 (2003); Y.~Sarazin and Ph.~Ghendrih,
Phys.\ Plasmas {\bf 5}, 4214 (1998).
%
\bibitem{ct:benkadda}
S.~Benkadda, X.~Garbet, and A.~Verga, Contib.\ Plasma Phys.\ {\bf 34},
247 (1994); O.~Pogutse, W.~Kerner, V.~Gribkov, S.~Bazdenkov, and
M.~Osipenko, Plasma Phys.\ Control.\ Fusion {\bf 36}, 1963 (1994).
%
\bibitem{ct:model}
A complete derivation of the model will be presented in an extended
report of these results. See however V.~Naulin, J.~Nycander
and J.~Juul Rasmussen, Phys.\ Rev.\ Lett.\ {\bf 81}, 4148 (1998) and
O.~E.~Garcia, J.~Plasma Phys.\ {\bf 65}, 81 (2001).
%
\bibitem{ct:garcia}
O.~E.~Garcia,  N.~H.~Bian, J.-V.~Paulsen, S.~Benkadda and K.~Rypdal,
Plasma Phys.\ Control.\ Fusion {\bf 45}, 919 (2003).
%
\bibitem{ct:gb}
O.~E.~Garcia and N.~H.~Bian, ``Bursting and large-scale intermittency
in turbulent convection with differential rotation'' accepted for
publication in Phys.\ Rev.~E (August 2003).
%
\bibitem{ct:naulin}
V.~Naulin, J.~Nycander and J.~Juul Rasmussen,
Phys.\ Plasmas {\bf 10}, 1075 (2003).
%
\bibitem{ct:s}
P.~C.~Stangeby, ``The plasma boundary of magnetic fusion devices''
(Bristol and Philidelphia: Institute of physics publishing, 2000).
%
\end{thebibliography}
\end{document}